# Electrical Response of Nanofluidic Systems Subjected to Viscosity Gradients

Ramadan Abu-Rjal[1], Zuzanna S. Siwy[2], and Yoav Green[1],*

[1] Department of Mechanical Engineering, Ben-Gurion University of the Negev, Beer-Sheva 8410501, Israel
[2] Department of Physics and Astronomy, University of California, Irvine, California 92697, United States

It is expected that the introduction of a viscosity gradient across a nanofluidic system will drastically vary its current-voltage response, $i$–$V$. However, to date, there is no self-consistent theoretical model that can be used to fully characterize such a system. This work provides an internally self-consistent model that details all the key characteristics of ion transport through a nanofluidic system for an arbitrary viscosity field. In particular, this work addresses three separate issues. First, we provide a new expression for the Ohmic conductance, $g_{Ohmic} = i/V$. Second, several previous theoretical studies have suggested that the introduction of a viscosity gradient can result in the shift of the $i$–$V$ such that it does not cross the origin. This work unequivocally shows that the $i$–$V$ is expected to always cross the origin. Third, we demonstrate that even without electroosmotic flows, the introduction of a viscosity gradient results in current rectification. Importantly, all theoretical results are verified by non-approximated numerical simulations. This work provides the appropriate framework to analyze and interpret experimental and numerical simulations of nanofluidic systems subject to a viscosity gradient.

## 1   Introduction

Permselectivity is a symmetry-breaking property allowing for the preferential passage of ions of a particular charge across a nanochannel. This property stems from the presence of surface charge on the nanochannel walls. Many materials (graphene and graphene-oxide membranes[1], carbon nanotubes[2] and carbon nanotubes porins[3], silicon nanochannels/nanopores[4,5], wood-based cellulose[6], single layer MoS2[7], boron nitride[8], and more; see the recent review[9,10] for a detailed list) that exhibit permselectivity can be found in various applications, such as desalination[11], energy harvesting[12], biomolecule sensing[13], and fluid-based electrical circuits[14,15]. In all of these applications, the ion-selective (charged) material [Region 2 in **Figure 1**(a)] is flanked by two uncharged regions filled with an electrolyte [Regions 1 and 3 in **Figure 1**(a)]. These two uncharged regions are commonly termed "diffusion layers" as the dynamics in these regions are dominated by the diffusion equation.

* Email: yoavgreen@bgu.ac.il
ORCID numbers:   R.A-R.: 0000-0002-1534-9710
                 Z.S.S: 0000-0003-2626-7873
                 Y.G.: 0000-0002-0809-6575

All these systems are characterized by the current-voltage ($i$–$V$) response. In general, the $i$–$V$ can be nonlinear and non-symmetric around the origin[14,16]. The nonlinearity is due to the nonlinearity of the governing equations, which couple ion transport with electrostatics. The asymmetry stems from internal asymmetries, such as variations in the surface charge – which yields unipolar[17,18] and bipolar systems[19–21] – or in the geometry[22,23], and external asymmetries, such as salt concentration[24], temperature[25], or viscosity gradients[26–29].

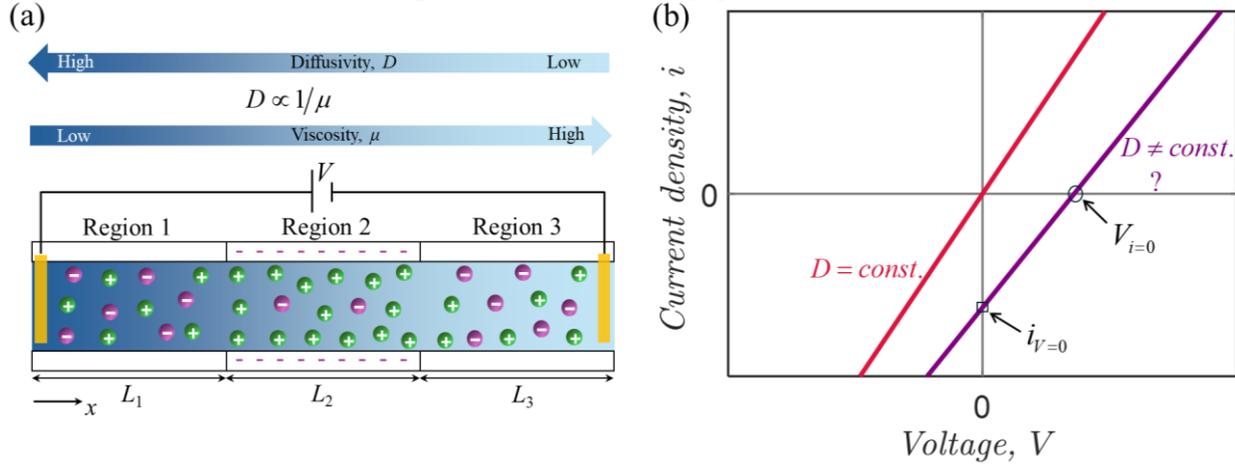

**Figure 1.** (a) Schematic of a one-dimensional three-layer system comprised of a permselective medium flanked by two diffusion layers containing a binary electrolyte with non-uniform diffusivities. The system is subject to a voltage drop $V$ (positive from left to right). The negative surface charge density results in an excess of positive counterions, represented by green spheres, over the negative coions, represented by purple spheres. (b) Schematic of the current density−voltage ($i$–$V$) response of the system with similar and different edge conditions. It has been suggested that a non-uniform diffusivity will result in the shift of the $i$–$V$ curve, resulting in two cross-over points – the zero-current voltage, $V_{i=0}$, and the zero-voltage current, $i_{V=0}$. Additionally, it is not known how the Ohmic conductance, $g_{Ohmic} = i/V$, varies with the diffusion gradient.

This work considers the general nonlinear $i$–$V$ response when the two reservoirs that flank the nanochannel of interest are filled with different media (i.e., different viscosities). However, for the sake of simplicity and brevity, in the introduction, we focus on the low-voltage-low-current linear Ohmic response and other key characteristics of ion transport in both symmetric and asymmetric systems. Additional attributes of the nonlinear $i$–$V$ response, which are not discussed in the introduction, are discussed in further sections.

The red line in **Figure 1**(b) depicts the low-voltage-low-current response of an "ideally" symmetric system. The term "ideally" refers to the fact that the edge conditions at the two ends of the system are the same (i.e., same bulk concentrations, same temperature, and more). For such an ideally symmetric system with symmetric geometries, the $i$–$V$ crosses through the origin, and the response is symmetric for all voltages.

In contrast, if the edge/external conditions are not the same, e.g., due to asymmetric concentrations or temperature gradients for energy harvesting via reverse-electrodialysis[24,30] and thermionics[25], respectively, then the $i$–$V$ is not only asymmetric but also shifted from the origin. The purple line in **Figure 1**(b) schematically depicts such a curve. Importantly, it can be observed that this line has two intercepts, which are the zero-current voltage, $V_{i=0}$, and the zero-voltage current, $i_{V=0}$. The values of $V_{i=0}$ and $i_{V=0}$ are strongly dependent on the mechanism leading to the shift (i.e., concentration gradient, temperature gradient), as well as the geometry and surface charge density of the system. In this work, we will focus on a viscosity gradient, and as such, we will not review the broad literature related to $V_{i=0}$ and $i_{V=0}$ for the other mechanisms. Instead, we will refer the interested readers to Refs. [24] and [25] for a review of $V_{i=0}$ and $i_{V=0}$ when the system is subjected to asymmetric bulk concentrations and asymmetric temperatures, respectively.

Two recent theoretical reports[31,32] have suggested that, similar to a concentration or temperature gradient, a viscosity gradient between the two ends of the system can also shift the $i-V$ curve such that finite $V_{i=0}$ and $i_{V=0}$ are observed. Both theoretical models evoked statistical mechanics concepts and included several embedded ansatzes into their model to create this shift. These reports, however, conflict with experimental findings[26–29], which measured several $i-V$'s of varying systems subjected to viscosity gradients and where the $i-V$ always appeared to cross the origin. In this work, using continuum mechanics, we provide theoretical support to the experiments, demonstrating that the $i-V$ is not expected to shift. We begin with describing our model (Sec. 2) and its results (Sec. 3). In Sec. 3.5 we will discuss how our models differ from those given in Refs.[31,32].

In addition to addressing the issue of whether the $i-V$ curve crosses the origin, this work also addresses two additional open questions. First, we discuss how the Ohmic conductance, $g_{Ohmic} = i/V$, varies if a diffusion coefficient gradient is introduced into the system. To this end, we have derived a novel expression for $g_{Ohmic}$ as a function of the diffusion coefficient gradient. Second, we have addressed the origin of current rectification. Past works have suggested that electroosmotic flow (EOF) is the responsible mechanism for current rectification in nanofluidics systems with viscosity gradients. Here, we show that convection-less ion transport can produce similar transport characteristics.

## 2   Model.

In order to describe the electrochemical properties of the system shown in **Figure 1**(a), we utilize a continuum mechanics approach – by solving the classical Poisson-Nernst-Planck equations that have been modified to account for viscosity gradients – or, more precisely, diffusion gradients. Our approach differs from the statistical mechanics approach of Refs.[31,32], because it is entirely self-consistent and does not require any ad-hoc assumptions, such as ignoring the effect of the nanochannel entrance and electro-migration contribution[31] or imposing periodic flux boundary conditions (where particles that would cross a boundary are counted and moved to the opposite boundary)[32]. Importantly, in contrast to the statistical physics models, our solution is derived analytically, eliminating the need for numerical evaluation to achieve a final result.

Details of the derivation are included in the Supporting Information. Here, we describe the fundamental physics concepts we utilized. First, the Stokes-Einstein relation

$$D = \frac{\Re T}{6\pi N_A R_H} \frac{1}{\mu},\qquad(1)$$

relates the local viscosity, $\mu$, to the local diffusion coefficient $D$. Here $N_A$ is the Avogadro constant, $R_H$ is the hydrodynamic radius (or the Stokes radius) of the ion, $\Re$ is the universal gas constant, and $T$ is the absolute temperature. So long as the temperature is uniform (which we shall assume here), $\mu$ and $D$ are inversely proportional to each other, such that our work, which focuses on gradients in the diffusion coefficients, is equivalent to those focused on viscosity gradients. Second, if $R_H$ for both ions are the same, then the diffusion coefficients for both ions are the same as well. For KCl, since $\tilde{R}_{H,K} \sim 1.25\,[\text{Å}]$ and $\tilde{R}_{H,Cl} \sim 1.21\,[\text{Å}]$, Eq. (1) implies that $D_K \cong D_{Cl}$. Thus, similar to the experimental works[26–29], we will take $D_K \equiv D_{Cl}$ and denote this generally as $D$. Third, while it is common to assume that $D$ is spatially independent throughout the system, we will consider the most general flux equation where $D$ depends on the axial position, $x$. Hence, our only "modification" of the Nernst-Planck equation is to remove the oversimplifying (and often correct) assumption of a uniform $D$, and to consider the more general $D(x)$. [Later, in Sec. 3.5, we will present this "modified" equation, Eq. (13), and discuss the role of $D(x)$]. Finally, in this work, we focused on the current density, $i$, and not the current, $I$, which are trivially related by $I = iA$, where $A$ is the cross-sectional area in 3D or the height in 2D. There are two reasons for this. While experiments measure $I$, this value is size-dependent, while $i$ is size-independent, making it a more general and robust parameter for comparing two different systems. Also, considering one-dimensional (1D) models drastically reduces the mathematical complexity involved and allows for the derivation of a simple and tractable equation. The differences between 1D and two-dimensional (2D) [or three-dimensional (3D)] are discussed in future research (Sec. 3.7.1)

In the Supporting Information, we have derived a general $i-V$ relation for a microchannel-nanochannel system, as shown in **Figure 1**(a), applicable to an arbitrary diffusion field $D(x)$. Each region has a length $L_k$, $k=1,2,3$. In contrast to Regions 1 and 3, which are uncharged, Region 2 has an excess counterion concentration $N$ (due to a negative surface charge), allowing for preferential transport of counterion (here, they are positive) over the (negative) coions. At the two ends of the system, we assume that the bulk concentrations are the same, $c_{bulk}$, and the system is under an applied potential drop of $V$, which leads to an electrical current $i$.

The resultant steady-state convection-less current-voltage response, $i-V$, is given by

$$V = V_{th}\left[\left(\frac{i}{Fj}+1\right)\ln\frac{\overline{c}_3}{\overline{c}_1} + \ln\frac{2\overline{S}_1+N}{2\overline{S}_3+N} + \frac{2\overline{S}_3 - 2\overline{S}_1 + j\eta_2}{N}\right]. \qquad (2)$$

Here, $V_{th} = \Re T/F$ is the thermal voltage, where $F$ is the Faraday constant. The voltage drop depends on several terms, including the interfacial concentrations $\overline{c}_1 = c_{bulk} - j\eta_1/2$ located at $x = L_1$ and $\overline{c}_3 = c_{bulk} + j\eta_3/2$ located at $x = L_1 + L_2$; $j$ is the salt current density [see Supporting Information and Refs.[17,18,20] for more information on $j = j_+ - j_-$ and $i = F(j_+ - j_-)$, note that the definitions for the positive and negative flux densities, $j_\pm$, are given below in Eq. (13)]; and three terms/functions

$$\eta_{k=1,2,3} = \int_{L_k}\frac{dx}{D_k(x)}, \qquad (3)$$

that account for the integrated effect of diffusion in all three layers (these $\eta_k$ terms will be discussed shortly). Finally, there are two additional functions, $\bar{S}_{m=1,3} = \sqrt{\tfrac{1}{4}N^2 + \bar{c}_m^2}$, that depend on the interfacial concentrations.

Several comments regarding Eq. (2) are warranted. First, the response is given in terms of $V(i)$ and not $i(V)$. While both are equivalent, and the $i-V$ is always plotted as $i(V)$, it is much more convenient to theoretically calculate $V(i)$ – this is because $V(i)$ can be given in a "compact form", whereas the $i(V)$ cannot be written in a "compact form" [note that $i(V)$ is a transcendental equation]. Second, the $i-V$ also depends on $j$. However, $i$ and $j$ are not two independent fluxes. Rather, they are two interdependent fluxes, with different units, that are determined by a single transcendental equation [i.e., $i(j)$] that is not given in the main text but can be found in the Supporting Information. Third, Eq. (2) depends both on the excess counterion concentration $N$ and the bulk concentration, $c_{bulk}$. Equation (2) holds for all values of $N$ and $c_{bulk}$ and thus holds for any selectivity (ideal, non-ideal, and vanishing). As will be shown later, the parameter that determines the selectivity is the ratio $N/c_{bulk}$. In particular, when $N/c_{bulk} \gg 1$, the system is ideally selective and when $N/c_{bulk} \ll 1$, the system is vanishingly selective. Fourth, this equation holds for any given diffusion field, $D(x)$ – how the $D(x)$ field is determined is beyond this work. Fifth, this expression is explicitly geometry-independent but implicitly dependent on the geometry through the $\eta_k$ terms [Eq. (3)]. Sixth, this equation was recently derived[17,18] for the case of uniform $D$, where the parameter $\eta_k = L_k/D$ represents the 1D resistance of each region. Here, the physical meaning is the same, except that resistance is integrated in each region.

## 3 Results and discussion

### 3.1 Numerical simulations

Throughout this work, we will compare our theoretical results to numerical simulations of the non-approximated Poisson-Nernst-Planck (PNP) equations that were conducted with the Transport of Diluted Species and Electrostatic modules in Comsol. Additional information on the numerical simulation methods and the selected parameter set can be found in Supporting Information Sec. 3 and **Table S1**, respectively.

Importantly, in contrast to theory, where $D(x)$ can be left arbitrary, in numerical simulations, they must be defined explicitly. Therefore, we must first present the considered scenarios, and the rationale for choosing them. Then, we will present their results. Our approach of considering (several) specific scenarios before generalizing for all cases is done in order to increase the intuition regarding the results.

### 3.2 Considered diffusion fields

**Figure 2** presents the four representative cases considered in this work on how $D(x)$ might behave (the mathematical details are given in **Table 1**):
- Case 1: $D(x)$ is uniform everywhere [**Figure 2**(a)].
- Case 2: the diffusion layers are well mixed such that the diffusion coefficient in Regions 1 and 3 are constant, and the linear change occurs only in Region 2 [**Figure 2** (b)].
- Case 3: the change of diffusion coefficient is linear along the axis between the two ends of the system [**Figure 2**(c)].

- Case 4: similar to Case 2 in that the diffusion coefficients in the diffusion layers are constants, but the diffusion coefficient in the charged region changes in a stepwise manner [**Figure 2**(d)]. One can think of this as an immiscible fluid meniscus whose thickness is substantially smaller than the length $L_2$. This scenario best matches the experimental setup of Ref.[29].

For the four scenarios given in **Table 1**, we use the following notations for the cumulative lengths: $\Delta_1 = L_1, \Delta_2 = \Delta_1 + L_2, \Delta_3 = \Delta_2 + L_3$. At $x = 0$ the diffusion coefficient is $D_l$ (short for $D_{left}$), while at $x = \Delta_3$, the diffusion coefficient is $D_r$ (short for $D_{right}$). In general, $D_l$ and $D_r$ can take any value, but here, $D_l$ is always that of a KCl electrolyte in an aqueous solution, and its diffusion is given by $D_{KCl}$, while the other bulk value can have any value, e.g. $D_r = \alpha D_l$, where $\alpha$ is the ratio of the two edge diffusion coefficients. Note that $\alpha \leq 1$, implies a higher viscosity.

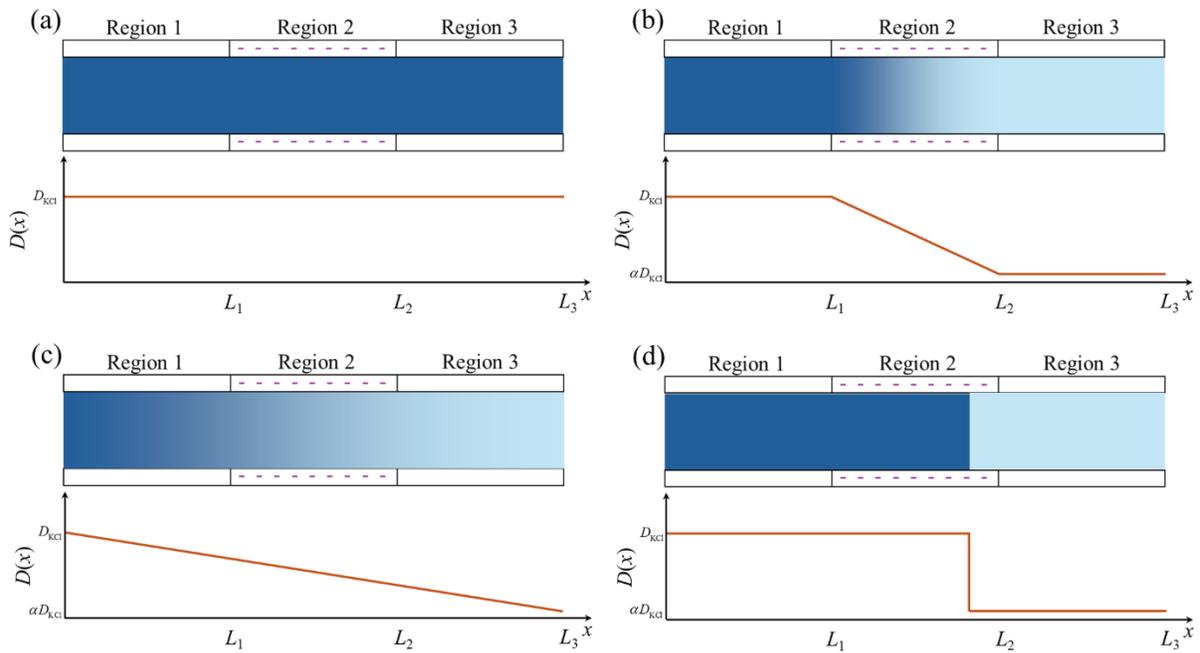

**Figure 2**. Schematics representation of the four considered diffusion fields: (a) constant diffusion field, (b) constant diffusion field in the diffusion layers with a linear drop in Region 2, (c) a linear drop of the diffusion across the two ends of the system, and (d) stepwise change in the diffusion within Region 2. See **Table 1** for the mathematical expressions of these four scenarios.

**Table 1**: The diffusivity distributions for the four cases considered in this manuscript. The diffusion coefficients $D_{k=1,2,3}(x)$ in the three regions in **Figure 1**(a) and **Figure 2** are given in the first columns, followed by the $\eta_{k=1,2,3}$ terms. For brevity, we denote $D_\delta = D_l - D_r$ and $\beta = \alpha - 1$. In Case 4, $H$ is the Heaviside step function at the point $x_0$ (see main text for the $x_0$ considered in this work).

| Case | $D_1(x)$ | $D_2(x)$ | $D_3(x)$ | $\eta_1$ | $\eta_2$ | $\eta_3$ |
|------|----------|----------|----------|----------|----------|----------|
| 1 | $D_{KCl}$ | $D_{KCl}$ | $D_{KCl}$ | $\dfrac{L_1}{D_{KCl}}$ | $\dfrac{L_2}{D_{KCl}}$ | $\dfrac{L_3}{D_{KCl}}$ |
| 2 | $D_l$ | $D_l + D_l\beta\dfrac{x-L_1}{L_2}$ | $D_r$ | $\dfrac{L_1}{D_l}$ | $-\dfrac{L_2}{D_\delta}\ln\alpha$ | $\dfrac{L_3}{D_{3b}}$ |
| 3 | $D_l\dfrac{\Delta_3+\beta x}{\Delta_3}$ | $D_l\dfrac{\Delta_3+\beta x}{\Delta_3}$ | $D_l\dfrac{\Delta_3+\beta x}{\Delta_3}$ | $\dfrac{\Delta_3}{D_\delta}\ln\dfrac{\Delta_3}{\Delta_3+\beta\Delta}$ | $\dfrac{\Delta_3}{D_\delta}\ln\dfrac{\Delta_3+\beta L}{L_3+\alpha\Delta}$ | $\dfrac{\Delta_3}{D_\delta}\ln\dfrac{L_3+\alpha\Delta_2}{\alpha\Delta_3}$ |
| 4 | $D_l$ | $D_l - D_\delta H(x-x_0)$ | $D_r$ | $\dfrac{L_1}{D_l}$ | $\eta_2^{(4)}$ a | $\dfrac{L_3}{D_r}$ |

a $\eta_2^{(4)} = \dfrac{D_\delta(L_1-x_0)}{D_l D_r} + \dfrac{L_2}{D_r}$

Several comments regarding Cases 2 and 4 are needed. Both represent feasible and physical fields. They are similar in that they have the same diffusion fields in the diffusion layers, and thus, they have the same $\eta_1$ and $\eta_3$ (**Table 1**). However, they are different in that the diffusion field in Region 2 is different such that $\eta_2$ is different (**Table 1**). However, when

$$x_0 = L_1 + \left(\dfrac{D_r}{D_\delta}\ln\alpha + 1\right)\dfrac{D_l}{D_\delta}L_2, \qquad (4)$$

$\eta_2$ of these two scenarios are the same. Then, one can expect that the resultant $i-V$ in Eq. (2) will be the same. In this work, we will consider the scenario that Eq. (4) holds. Then, we will show that the $i-V$ of Cases 2 and 4 are identical. However, as we will also show, the concentration and electric potential distribution associated with each field will be different.

### 3.3 Current-voltage response

**Figure 3**(a) demonstrates the remarkable correspondence between the theoretical and numerically calculated $i-V$'s for all four scenarios. Several general observations are immediate:
- The $i-V$'s are not linear (discussed further in Sec. 3.3.1).
- When $D(x)$ in non-uniform, the $i-V$ is asymmetric, such that the current is rectified (discussed further in Sec. 3.3.2). Importantly, the current is rectified without accounting for the effects of electroosmosis (discussed further in Sec. 3.3.3).
- The $i-V$ always crosses the origin such that $i_{V=0} = 0$ and $V_{i=0} = 0$. In Sec. 3.5, we will show that this result holds for arbitrary $D(x)$ such that $i_{V=0} \triangleq 0$ and $V_{i=0} \triangleq 0$.
- As predicted, Cases 2 and 4 have identical $i-V$.

### 3.3.1 Nonlinear $i-V$'s

Returning to Eq. (2), it can be noted that, in general, the $i-V$ are not linearly dependent. In fact, the second term in Eq. (2) includes a logarithmic dependence on the interfacial concentrations. As these concentrations change and the effect of depletion and enrichment (i.e., concentration polarization effect[33]) becomes more prominent, the voltage increases logarithmically. Alternatively, as the voltage is increased, a limiting current appears. It can be shown (see the Supporting Information) that the values of the limiting salt currents are given by

$$j_{\lim,V>0} = 2c_{bulk}/\eta_1, \quad j_{\lim,V<0} = -2c_{bulk}/\eta_3. \tag{5}$$

Note that given the $i(j)$ relation, one can then calculate the limiting electrical current, $i_{\lim}$. Most importantly, Eq. (5) shows that if the geometries are asymmetric, or viscosity gradient is present, or both, then $|j_{\lim,V>0}| \neq |j_{\lim,V<0}|$. Naturally, as a result, the electrical current response is asymmetric, where the larger current magnitude appears on the side with the lower value of $\eta$.

Naturally, and quite intuitively, it can be observed in **Figure 3**(a) that the limiting current is strongly voltage-dependent. This, too, can be expected since a larger viscosity (or lower diffusion) field would require a larger voltage drop in order to deplete the interface. In particular, observe that the inset of **Figure 3**(a) demonstrates that Cases 2 and 4 have limiting currents that appear at substantially larger voltages. This is because these two cases have the largest degree of asymmetry – and this is also why their positive and negative limiting currents have the largest ratio, **Figure 3**(b).

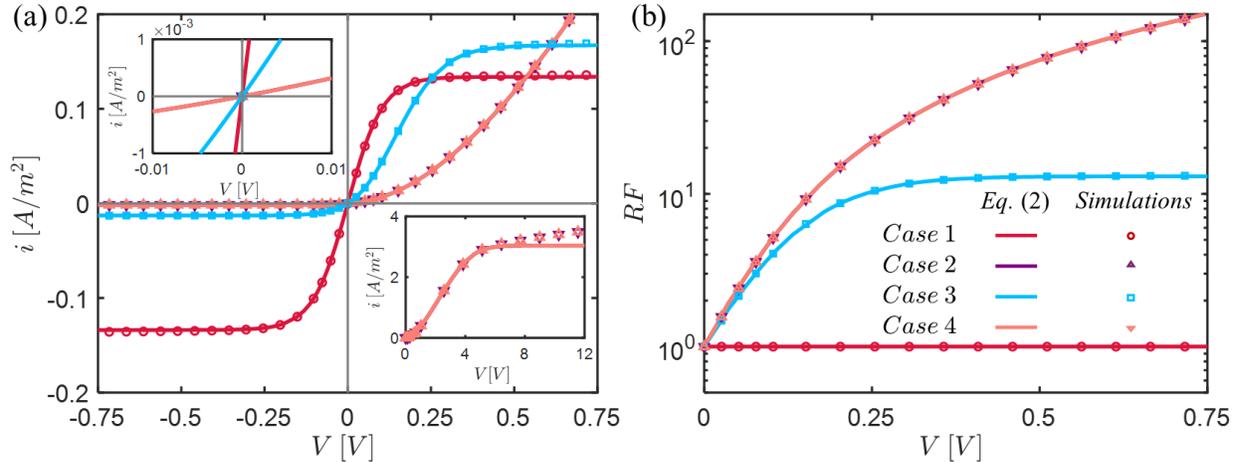

**Figure 3**. (a) Current density-voltage ($i-V$) response curves and (b) a semilog plot of the rectification factor ($\log_{10} RF$) for a highly permselective system ($N/c_{bulk} = 100$ with $c_{bulk} = 0.1 [\text{mol/m}^3]$) for the various diffusivity distributions given in **Table 1** when $L_1 = L_2 = L_3$. Lines represent values calculated from Eq. (2), while the symbols represent the non-approximated numerical simulations (additional simulation parameters are provided in **Table S1** in the Supporting Information). (a-Left top inset): A zoomed view of the $i-V$ curve near the origin, showing that the $i-V$ always crosses the origin. (a-Right bottom inset): The higher positive voltages $i-V$ for Cases 2 and 4.

One last comment is needed regarding the $i-V$ curves shown here. Note, the $i-V$ are plotted based on Eq. (2), which accounts for the microchannels/diffusion layers. Removal of the diffusion layers, which can be achieved by inserting $\eta_{k=1,3} = 0$ into Eq. (2), leads to several relations, $\bar{c}_1 = \bar{c}_3$, $\bar{S}_1 = \bar{S}_3$, and infinite limiting currents [Eqs. (5)]. Additionally, the response between the voltage drop and the salt current density, $V = V_{th} j\eta_2 / N$, becomes linear. Soon, we will transform $j$ into $i$, and characterize the linear Ohmic conductance, $g$ [Eq. (16)], which predicts that for a nanochannel-only system, the response is always linear and given by $i = gV$. In contrast, we will show that for systems that account for the diffusion layers, the relation $i = gV$, holds only for low voltages. The Ohmic conductance is discussed further below in Sec. 3.6.

### 3.3.2 Current rectification

Since the limiting currents, which depend on the geometry and diffusion fields, are different for different polarities, then naturally, the $i-V$ curve is asymmetric, leading to current rectification. The asymmetry of the $i-V$ is often characterized by the rectification factor

$$RF = \left|\frac{i_{V>0}}{i_{V<0}}\right|. \tag{6}$$

The rectification factor for all curves shown in **Figure 3**(a) is shown in **Figure 3**(b). As expected, the case with the largest integrated asymmetry in the diffusion coefficient in Regions 1 and 3 (Cases 2 and 4) exhibited the largest rectification, which increases with the increase of voltage, while Case 1 produced $RF = 1$. Note that here, to emphasize the importance of $D(x)$, we have considered symmetric geometries. One can further increase the observed rectification by breaking the geometric symmetry. (The $i-V$ and $RF$ for Case 1 with asymmetric geometries is shown in Ref. [22]).

### 3.3.3 Convection-less rectification

It is worthwhile to note that the mechanism responsible for the rectification in cases considered here differs drastically from those used to explain the experimentally measured rectification reported in Refs. [26–28]. The geometry in all three works was different: Ref. [26] used conical nano- and micropores, Ref. [27] used a cylindrical micropore, and Ref.[28] used a cylindrical nanopore. However, regardless of the geometry, the experimental conditions in all of these works corresponded to what we term the non-selective limit $N \ll c_{bulk}$. Here, it was thought that (the low) selectivity should not play a key role in rectification. Therefore, it was suggested that the rectification was due to geometric asymmetry combined with electroosmotic flow (EOF) effects.

In the main text, and thus far, we have primarily focused on the $N \gg c_{bulk}$ scenario, which shows remarkably strong rectification. The proposed mechanism here does not evoke convection. However, while the change from $N \gg c_{bulk}$ to $N \ll c_{bulk}$ changes the selectivity of the system, which results in both a qualitative (highly versus vanishingly selective) and quantitative change in the $i-V$ and $RF$, it does not change the fact that there is still an asymmetry in the response due to the gradient in $D(x)$. Supplemental **Figure S2** provides a complementary figure to **Figure 3** (which was plotted for $N \gg c_{bulk}$) for the non-ideal case of $N = c_{bulk}$, showing that there is also a non-unity rectification even though EOF is not considered here. We do not doubt that EOF would only enhance the rectification – but the effects of EOF are left for future work.

We all note that the $N \ll c_{bulk}$ scenario will also have rectification (**Figure S3**). However, as the $N / c_{bulk}$ becomes smaller, the symmetry breaking becomes weaker and will only appear at higher voltages. The reason for the change in the behavior of the system is due to the change in the selectivity of the system, which is characterized by the transport number (discussed in Sec. 3.5). For $N \gg c_{bulk}$, corresponding to ideal selectivity, there is a strong asymmetry in the positive and negative ion response, which, combined with $D(x)$, leads to asymmetry in the $i-V$. For $N \ll c_{bulk}$, there is virtually no asymmetry between positive and negative ions, such that the only asymmetry in the system can be attributed to $D(x)$.

### 3.4 Concentration and electric potential distributions

**Figure 4** plots the concentration of negative ions and electric potential distributions for Cases 1-4 for both positive and negative applied voltages for an ideally selective nanofluidic system [**Figure 1**(a)]. Here, we have only plotted the simulation results. We do not show the theoretical predictions, which show remarkable overlap to the simulations, for two reasons: clarity and the theory does not capture the sharp changes due to the Donnan potential that occur with one Debye length at the interfaces located at $x = \Delta_1$ and $x = \Delta_2$. Also, here, we have not shown the positive concentration distribution for two reasons: within the diffusion layers (i.e., Regions 1 and 3), the positive and negative concentrations are identical and indistinguishable, while in the nanochannel (i.e., Region 2) the Donnan potential results in a substantial jump of the positive concentration [such that the distribution is not within the range of concentrations shown in **Figure 4**(a-b)].

Unsurprisingly, it can be observed that the distributions change with the diffusion coefficient fields and the $\eta$-functions. In particular, let us focus on Cases 2 and 4. **Table 1** shows that $\eta_1$ and $\eta_3$ are identical [regardless of Eq. (4)] – thus, the concentrations and potentials in Regions 1 and 3 are the same. Now, even when Eq. (4) holds, and the potential distributions in Region 2 are the same [pink and purple lines in **Figure 4**(c-d) are in perfect overlap] – this is what leads to the identical $i-V$'s – the concentration distribution in Region 2 is different [pink and purple lines in **Figure 4**(a-b)].

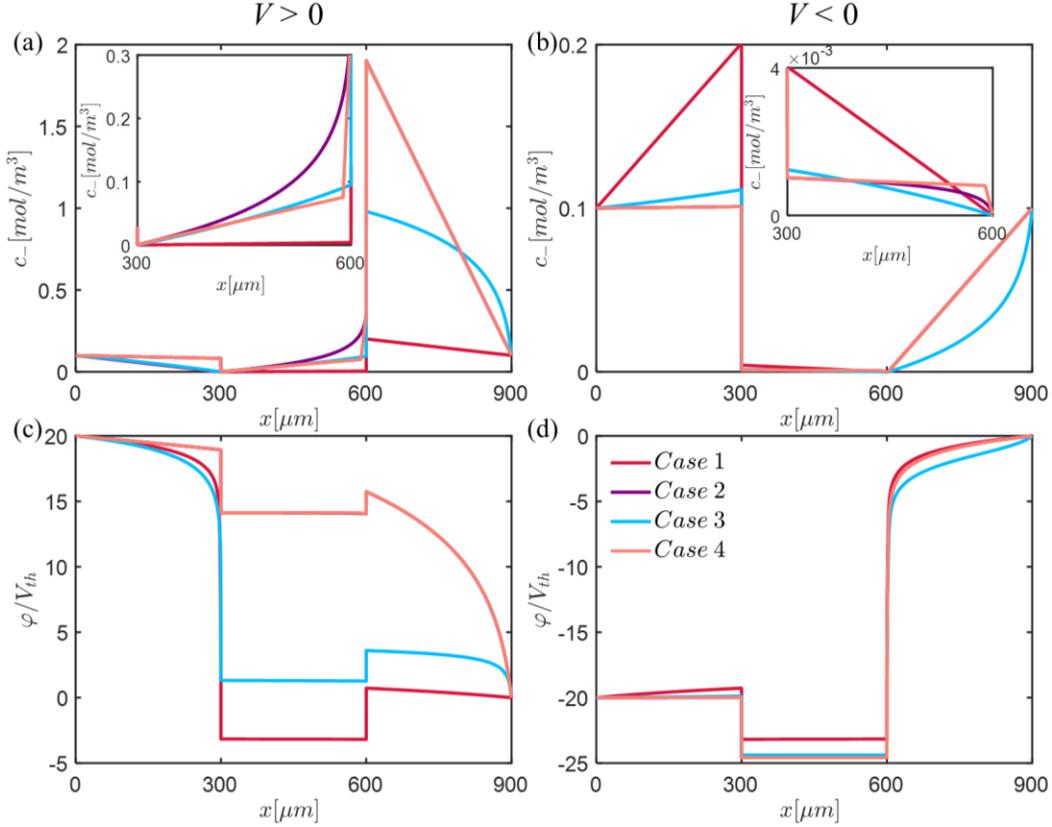

**Figure 4**. (a and b) Anion concentration $c_-$ and (c and d) potential $\varphi$ distributions for positive (a and c) and negative (b and d) voltages ($V = \pm 20V_{th}$) for a highly permselective system ($N/c_{bulk} = 100$ with $c_{bulk} = 0.1[\text{mol}/m^3]$) for the various diffusivity distributions given in **Table 1** when $L_1 = L_2 = L_3$. The insets in (a) and (b) are a zoomed view of the concentration in the permselective region. Additional simulation parameters are given in **Table S1**.

### 3.5 The $i-V$ intercepts: $i_{V=0} = 0$ and $V_{i=0} = 0$

To provide evidence that Eq. (2) always passes through the origin, we leverage that $i$ and $j$ are related through the (non-dimensional) transport number (for the positive species) $\tau = \frac{1}{2}$

$$\tau = (1 + Fj/i)/2. \tag{7}$$

The transport number varies between $\frac{1}{2}$ and 1. When the system is vanishingly selective, such that $N/c_{bulk} \ll 1$, the contribution of both species is identical such that $\tau = \frac{1}{2}$. In contrast, when the system is ideally selective, such that $N/c_{bulk} \gg 1$, only the positive species is transported. Then, $i = Fj = Fj_+$ and the transport number is unity, $\tau = 1$. The transition from $\tau = \frac{1}{2}$ to $\tau = 1$ will be demonstrated in the next sub-section (Sec. 3.6).

At low currents, this relation is given by (see Supporting Information for the derivation)

$$\tau = \frac{1}{2} + \frac{N}{4c_{bulk}} \frac{1}{S/c_{bulk} + (\eta_1 + \eta_3)/\eta_2}, \tag{8}$$

where $S = \sqrt{\frac{1}{4}N^2 + c_{bulk}^2}$. Note here that $S$ differs from $\bar{S}_{1,3}$ in Eq. (2). While both have the same form, $S$ depends on $c_{bulk}$, whereas $\bar{S}_{1,3}$ depend on the interfacial concentration $\bar{c}_{1,3}$.

Inserting Eq. (8) into Eq. (2) and taking the Taylor series for $i \ll 1$ yields

$$V_{i \to 0} = \frac{\Re T}{F^2} \left\{ [2\tau - 1]\frac{\eta_2}{N} + \left[\frac{1}{2} + (2\tau - 1)\frac{S}{N}\right] \frac{\eta_1 + \eta_3}{c_{bulk}} \right\} i. \quad (9)$$

This equation can be rewritten as

$$V_{i \to 0} = ri, \quad (10)$$

$$r = g^{-1} = \frac{\Re T}{F^2} \left\{ [2\tau - 1]\frac{\eta_2}{N} + \left[\frac{1}{2} + (2\tau - 1)\frac{S}{N}\right] \frac{\eta_1 + \eta_3}{c_{bulk}} \right\}, \quad (11)$$

where Eq. (10) is Ohm's law with $r$ having a meaning of "resistance density" (i.e., the total resistance, $R$, multiplied by the uniform cross-section area of the system, $A$, such that $r = RA$).

Before we proceed with our analysis of the resistance (or conductance), it is important to observe that the simple form of Eq. (9) implies a rather remarkable result. Regardless of the microscopic details of the diffusion field, the $i-V$ always crosses the origin such that $i_{V=0} \triangleq 0$ and $V_{i=0} \triangleq 0$. In retrospect, this result should not be surprising. The Nernst-Planck equations are the gradients of the electrochemical potentials. The electrochemical potential is given by

$$\mu_\pm = \Re T \ln c_\pm + z_\pm F \varphi + \mu_{ref,\pm}. \quad (12)$$

Then, the fluxes are given by

$$j_\pm = -\frac{D(x)}{\Re T} c_\pm \mu_{\pm,x} = -D(x)\left[ c_{\pm,x} + \frac{z_\pm F}{\Re T} c_\pm \varphi_{,x} \right], \quad (13)$$

where the comma denotes a derivative relative to $x$. In the 1D steady-state scenario, the Nernst Planck equations are $j_{\pm,x} = 0$. One is tempted to look at Eq. (13), where it appears that there is an explicit dependency on $D(x)$ yielding the expectation that the fluxes, and therefore, the currents ($j$ and $i$) are diffusion gradient dependent. However, one should remember that the fluxes are the derivatives (or gradients) of the electrochemical potential $\mu_\pm$. This potential, given by Eq. (12), is independent of $D(x)$. Thus, a gradient in $D(x)$ does not induce a gradient in $\mu_\pm$, such that it does not induce a flux $j_\pm$. If the fluxes are not induced, then neither is an electrical current induced. Therefore, regardless of the imposed viscosity gradient, the $i-V$ must cross the origin. This simple explanation is entirely consistent with the mathematical result shown in Eq. (10) [and shown in **Figure 3**(a)].

We can now return to the difference between our continuum model and the statistical physical models of Refs.[31,32]. Note that the concept of the electrochemical potential is a statistical physical concept. Thus, in principle, our model and those previously reported[31,32] should give the same result. The question, therefore, is what are the fundamental differences between our and the earlier approaches that lead to different results and conclusions. The answer is remarkably simple.

In both Refs.[31,32], only the diffusion gradient term in Eq. (13) was considered together with the cumulative effects of the viscosity gradient on the motion of the particle. This approach, however, is equivalent to removing the electric potential from both Eqs. (12) and (13), and thus removing the coupling and interactions between all ions (ions of the same sign and of opposite sign). Mathematically, this is equivalent to ignoring the Poisson equation for the electric potential. Consequently, the models of Refs.[31,32] did not account for the fact that an electrical field will form when charge separation (due to the selectivity) occurs. One can then view those models to be single-particle models (in fact, Ref.[32] was a modified Langevin model for a single particle that does not interact with additional particles). In contrast, our continuum model is rooted on the Poisson equation and the resultant multi-body interactions. Our multi-particle model self-consistently accounts for all electric interactions of particles of the same species and particles of opposite species.

### 3.6 Ohmic conductance

We shall now return to the low-voltage-low-current response given by Eq. (11) and note that the resistance can be written in terms of its reciprocal, the Ohmic conductance density, $g = r^{-1}$. Here, $g$ is related to the Ohmic conductance, $G$, by $G = gA$ (while $R = r/A$). Since we are interested in an arbitrary cross-section, in the remainder, we will consider $g$.

As it turns out, Eq. (11) can be used to recapitulate several known results for the case of a constant $D$, as well as generalizing them for the case of $D(x)$. To that end, we shall demonstrate how Eq. (11) reproduces the well-known expression for the Ohmic conductance for constant $D$ [2,34–36]

$$g_{nano} = 2\frac{DF^2}{\Re TL_2}S = 2\frac{DF^2}{\Re TL_2}\sqrt{\tfrac{1}{4}N^2 + c_{bulk}^2}. \tag{14}$$

This expression is derived with an underlying hidden assumption that the effects of the diffusion layers are negligible ($\eta_1 = \eta_3 = 0$) and that the diffusion coefficient is constant [$D(x) = D$]. We now relax both assumptions, one at a time and in different order.

First, consider the case of the one-layer 'nanochannel' system (i.e., no diffusion layers such that $\eta_1 = \eta_3 = 0$), Eqs. (8) and (11) reduces to

$$\tau = \frac{1}{2} + \frac{N}{4S}, \tag{15}$$

$$g = g_{nano} = 2\frac{F^2}{\Re T}\frac{S}{\eta_2}. \tag{16}$$

Equation (16) is strikingly similar to Eq. (14) with one change: Eq. (16) depends on the more general $\eta_2$ term instead of on $D$ as in Eq. (14). Naturally, when $D(x) = D$ such that $\eta_2 = L_2/D$, we find that Eqs. (14) and (16) are identical. The dashed lines in **Figure 5**(b) show that the nanochannel-only conductance varies as the viscosity gradient is varied, where the uniform diffusion field has the expected highest conductance.

We now relax the assumption of negligible diffusion layer effects, such that $\eta_{1,3} \neq 0$, for the two extreme cases of ideal and vanishing permselectivity ($N \gg c_{bulk}$ and $N \ll c_{bulk}$). It can be shown that in these limits, the transport number and the conductance are given by (see Supporting Information)

$$\tau = 1 \quad , \quad g_{N \gg c_{bulk}} = \frac{F^2 c_{bulk}}{\Re T}\frac{N}{c_{bulk}\eta_2 + N(\eta_1 + \eta_3)}, \tag{17}$$

$$\tau \simeq \frac{1}{2} + O(N/c_{bulk}) \quad , \quad g_{N \ll c_{bulk}} = \frac{F^2 c_{bulk}}{\Re T} \frac{2}{\eta_1 + \eta_2 + \eta_3} . \tag{18}$$

Note that for the case $D(x) = D$, $\eta_k = L_k / D$ recapitulates the previous results of Ref.[17,18] (see these works for a detailed discussion).

**Figure 5** plots $\tau$ and $g$ for three scenarios:
- The nanochannel-only scenario ($L_1 = L_3 = 0$).
- A three-layered system when all lengths are the same, $L_1 = L_2 = L_3$, shown in **Figure 5**(a-b).
- A three-layered system when $L_1 = L_3 = L_2/100$, shown in **Figure 5**(c-d)

The nanochannel-only case is given as a reference for comparison relative to both three-layer system scenarios considered. The purpose of the two three-layered scenarios is to highlight the dependency of $\tau$ and $g$ on $D(x)$ and the geometry.

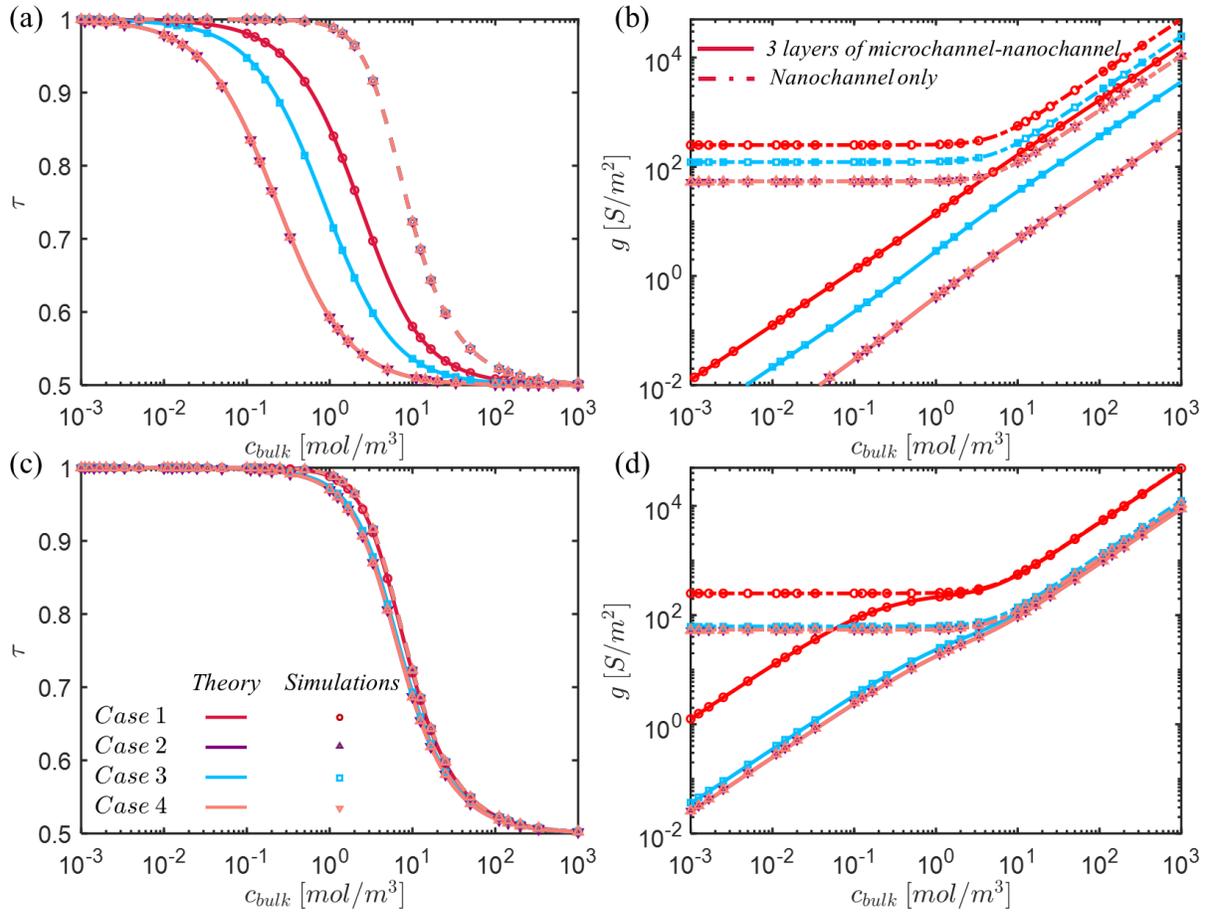

**Figure 5**. (a and c) Semilog$_{10}$ plots of the transport number and (b and d) log$_{10}$-log$_{10}$ plots of the low-voltage-low-current Ohmic conductance density versus the bulk concentration $c_{bulk}$ for the various diffusivity distributions given in **Table 1** and for two scenarios: (Top row) when all lengths are the same ($L_1 = L_2 = L_3$), and (Bottom row) when $L_{1,3} = L_2/100$. Dashed lines represent the transport number and conductance of the one-layer 'nanochannel' system [Eqs. (14) and (15)]. Solid lines represent the theoretically calculated values for the three-layer (micro-nano-micro channels) system [Eqs. (8) and Eq. (11)], while the symbols represent the non-approximated numerical simulations (simulation parameters are provided in **Table S1** in the Supporting Information).

**Figure 5**(a-b) shows the transport number and conductance for $L_1 = L_2 = L_3$. We note first that the transport number for all 4 cases in **Figure 5**(a) of the nanochannel-only system ($\eta_{1,3} = 0$) are the same regardless of $D(x)$. This is because Eq. (15) is independent of $D(x)$. Once the effects of the diffusion layers are accounted for such that $\eta_{1,3} \neq 0$, as can be expected, the transport number, given by Eq. (8), varies with $D(x)$. It can be observed that the effects of the adjoining microchannels (Regions 1 and 3) shift the response relative to the nanochannel-only system, where it can be observed that as the effects of the diffusion gradient become more dominant, the curve is further shifted to the left. This change is due to the fact that the existence of diffusion layers reduces the selectivity of the system (as shown in Ref. [17]) combined with the effects of $D(x)$. Smaller diffusion coefficients amplify the role of the diffusion layers. However, the change is not only in the shifting of the transition region to the left. It can also be observed with the "thickness" of the transition region from 1 (ideal selectivity) to ½ (vanishing selectivity). It can be observed that this "thickness" varies as the diffusion field varies.

**Figure 5**(b) shows several interesting results. First, even though the transport number of the nanochannel-only system is independent of $D(x)$, the conductance is not [Eq. (16)]. It can be observed that the modified nanochannel-only conductance varies as expected with $D(x)$ [with the uniform diffusion having the largest conductance]. Second, the conductance of a system with three layers does not saturate to a constant value as predicted by the nanochannel-only system [Eq. (16)]. Rather, it continues to decrease. This is because the resistances associated with the diffusion layers are non-negligible relative to that of the nanochannel (a similar result for uniform diffusion fields was shown in past works[18,22,37,38]). Intuitively, one expects to see a more drastic change when transitioning from vanishing to ideal selectivity (or vice versa), whereby the nanochannel resistance should become dominant. However, in this 1D $L_1 = L_2 = L_3$ scenario, the diffusion layer resistances are always of the same order of magnitude as the nanochannel resistance, or even larger. This is related to the third observation. At high concentrations, the slope is always unity. Yet, the curves are shifted with a strong dependence on the interplay of $\eta_2$ to $\eta_1$ and $\eta_3$ .[Eq. (11)]. As the resistances of the diffusion layers become larger, the curves shift downward.

To better simulate a 2D or 3D scenario where the nanochannel resistance is more dominant, we also consider the case where the nanochannel length is substantially larger than the diffusion layer lengths ($L_1 = L_3 = L_2/100$). Now, it can be expected that the permselective region will influence the transport characteristics of the system in a more dominant manner than in the case considered above. Indeed, **Figure 5**(c) shows that the transport number becomes less sensitive to the changes – in fact, the three-layer solution almost completely overlaps with the nanochannel-only solution. Combined **Figure 5**(a) and **Figure 5**(c) provide a clear indication that the transport number is highly dependent on the geometry and $D(x)$.

**Figure 5**(d) shows how the conductance depends on bulk concentration for the four cases of $D(x)$ considered here. Note that when the diffusion layers are significantly shorter than the permselective region, for Cases 2-4, the high concentration response is virtually identical to the nanochannel-only response [Eq. (16)]. However, this is expected since when the channel is not permselective, the effects of the diffusion layers are mostly negligible. The conductance at low concentrations, on the other hand, differs from Eq. (16), confirming that the diffusion layers

significantly affect ionic transport due to increased permselectivity. To explain this, we will analyze all cases through the lens of Case 1.

The behavior of Case 1, for the uniform diffusion field scenario, has been covered extensively in our past works.[18,22,37]. Here, we review the insights needed to understand the behavior of conductance of Cases 2-4. For all cases, the conductance of the three-layered system is plotted using Eq. (11). In all scenarios, at $N \ll c_{bulk}$, Eq. (11) is reduced to Eq. (18). In this circumstance, we are dominated by the $\eta_2$ term, and conductance approaches the nanochannel-only conductance [Eq. (16)] at high concentrations. At intermediate concentrations, $N = c_{bulk}$, for Case 1, the nanochannel response, which is now dominated by the effects of the surface charge $\sigma$, is, at the very least, the same order of magnitude as the remaining resistances, such that the response tries to saturate to a given value. This is the "transition region" where the redline's slope is almost zero. Then, at low concentrations, $N \gg c_{bulk}$, Eq. (11) reduces to Eq. (17), which can be further reduced to be

$$\tau = 1 \quad , \quad g_{N \gg c_{bulk}} \approx \frac{F^2 c_{bulk}}{\Re T (\eta_1 + \eta_3)} . \tag{19}$$

The response is now determined only by the diffusion layers, where the effects of the nanochannel are negligible. Cases 2-4 behave similarly to Case 1, with one notable difference: the "length" of the "transition region" is shortened. This is because the region in which the term is associated with the surface conductance [Eq. (14)], $g_{nano} \approx (DF^2 N)/(\Re T L_2)$, is now modulated to be $g_{nano} \approx (F^2 N)/(\Re T \eta_2)$. However, since $g_{nano}$ decreases, the transition region, in which it can change, is also decreased. Thus, the transition from Eq. (18), at high concentrations, to Eq. (17), at low concentrations, is shortened.

## 3.7 Future directions and implications
### 3.7.1 2D and 3D systems.

In this work, we have considered a purely 1D system whereby the cross-sectional area of the diffusion layers is equal to that of the nanochannel. Such a system represents a nanoporous membrane sandwiched between two microchannels rather than an individual nanochannel between two microchannels. However, except for a change in geometry, both systems are modeled in the same manner. Namely, they are governed by the same Poisson-Nernst-Planck equations. Typically, in both scenarios, cross-sectional 1D averages are used in the nanochannel/membrane/permselective material. The main difference lies in modeling the 2D or 3D effects of the diffusion layers/microchannels, whereby the diffusion layers have much larger cross-section areas than the permselective region. Thus, the system can no longer be modeled in a 1D manner, leading to one major mathematical change.

In 1D, the governing equation for the concentration in the diffusion layer, $c = c_+ = c_-$, is rather simple. It is $j = -2D(x)\partial_x c$, which can be solved for an arbitrary $D(x)$ [see Supplemental Information]. However, in a 2D or 3D system, the derivative with respect to $x$ in Eq. (13) needs to be replaced by the nabla/del ($\nabla$) operator (the vector form of the PNP equations are given in Sec. 3.2 in the Supporting Information) and $D(x)$ needs to be replaced by the more general $D(\mathbf{x})$. Then, the governing equation is modified such that $\nabla \cdot \mathbf{j} = -\nabla \cdot [2D(\mathbf{x})\nabla c] = 0$. For arbitrary $D(\mathbf{x})$, this equation cannot be solved analytically in a simple, straightforward manner such that one can derive a simple contractable expression such as the one given in Eq. (2). Thus, in 2D (or 3D), we no longer expect the $\eta_k$ terms to be given by a simple equation such as Eq. (3). Rather, we expect these $\eta_k$ terms to have a much more complicated expression (which depends on the 2D/3D geometry). How these $\eta_k$ behave will need to be addressed in future works. However, we will now demonstrate that the transition from 1D to 2D does not change the qualitative nature of the results in the work.

Before showing these results, two preceding comments are warranted. First, in scenarios involving a uniform diffusion coefficient with an asymmetric geometry[22,39,40], we know that the current is rectified, but the current crosses through the origin. In other words, geometry alone does not qualitatively change the results. Second, the entire argument regarding the electrochemical potential (given in Sec.3.5) is independent of a specific geometry [and is independent of whether the derivative is a 1D partial derivative or the nabla/del ($\nabla$) operator]. Combined, theoretical considerations suggest that a qualitative change does not take place.

**Figure 6**(a) shows the $I-V$'s (here $I=ih$, where $h$ is the height of the nanochannel; the complete geometric details are given in Sec. 3.2 in the Supporting Information) of the four cases considered in **Table 1**. We note that, as expected, all the $I-V$'s cross through the origin. We also note that the rectification factor of Cases 2 and 4 are still the largest [**Figure 6**(b)]. In 2D, the relatively large resistance of the nanochannel results in the transport number being dominated by the nanochannel (similar to the nanochannel-only result shown in **Figure 5**), and most of the curves are essentially independent of $D(x)$ [**Figure 6**(c)]. Finally, in **Figure 6**(d), we observe that the low-concentration conductance doesn't saturate to a zero slope. Rather, the slope for all curves at these low concentrations is one. The origin of this slope is the contribution of the microchannel resistances as well as the field-focusing resistances. Both contributions are well documented in our past works[18,22,37,38]. In contrast to our past works, here, the value of the conductance itself does depend on the $D(x)$. Once more, the details of how $\eta_k$ changes in 2D (or 3D) needs to be examined in future works.

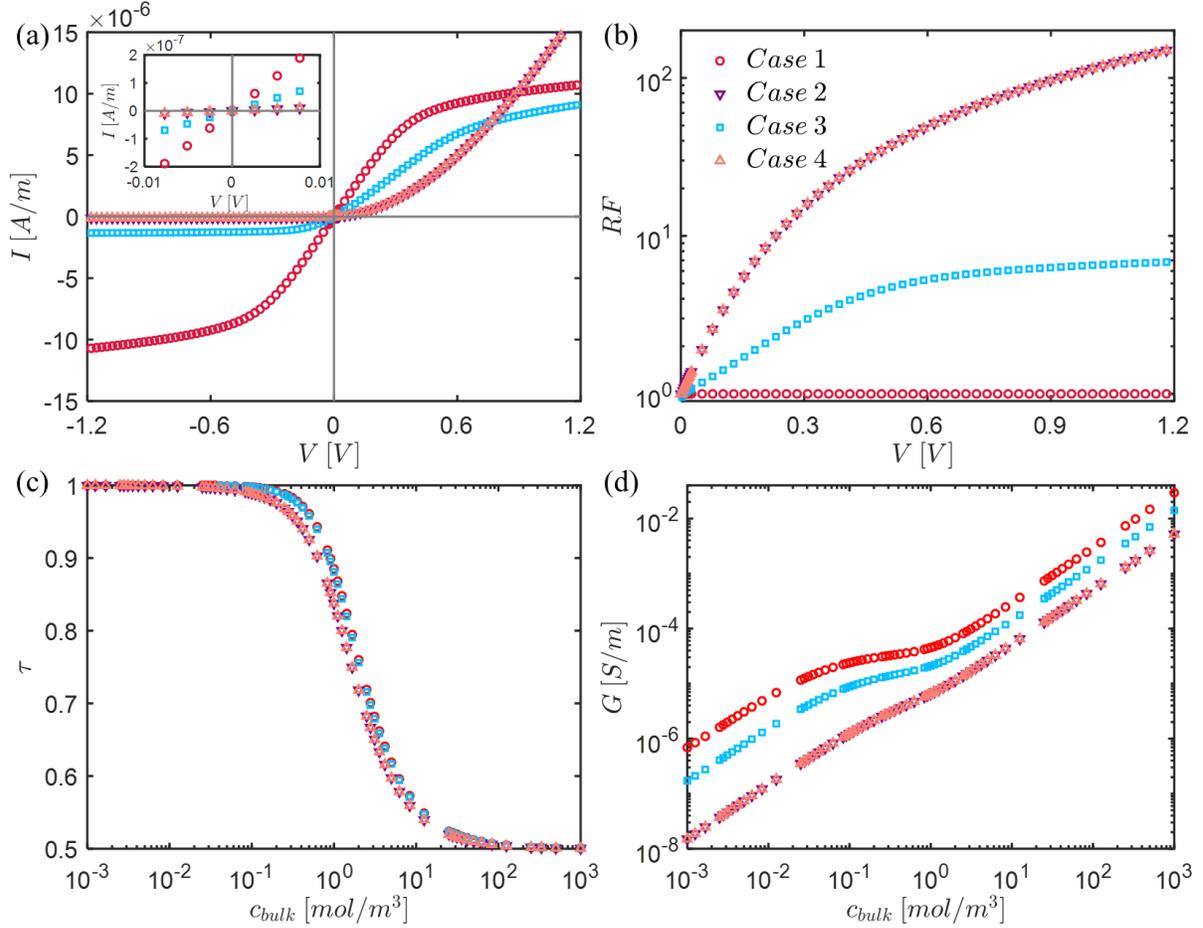

**Figure 6**. Simulation results of the 2D three-layer (micro-nano-micro channels) system for the various diffusivity distributions are given in **Table 1**. (a) The current-voltage ($I-V$) response curves for a highly selective system ($\tilde{N}/\tilde{c}_{bulk}=25$ with $c_{bulk}=0.1[\text{mol/m}^3]$). Inset: A zoomed view of the $I-V$ curve near the origin, showing that the $I-V$ always crosses the origin. (b) A semilog plot of the rectification factor ($\log_{10} \text{RF}$). (c) A semilog plot of the transport number versus the bulk concentration $c_{bulk}$ ($\log_{10} c_{bulk}$). (d) $\log_{10}$-$\log_{10}$ plots of the low-voltage-low-current Ohmic conductance density versus the bulk concentration. Simulation parameters for the geometry, portrayed in **Figure S1**, are provided in **Table S2** in the Supporting Information.

### 3.7.2 Physiological systems

This work also has implications for ion transport in physiological systems, such as transport across cell membranes[41–44]. In such systems, ions are transported from the inner part of the cell to the outer part or vice versa. Based on physiological conditions, the intercellular fluid and the extracellular fluids, and their mechanical properties (i.e., viscosity) can be very different, which in turn can lead to vastly different electrical properties (conductance, rectification, etc.'). While there remain many open questions regarding ion transport in physiological systems, this work can be used for the interpretation of experiments and numerical simulations.

### 3.7.3 Interplay of diffusion gradients with EOF

As already mentioned in Sec. 3.3.3, the effect of EOF on the $i-V$ response is still not understood. Future works will need to address how the addition of EOF will enhance the current and the asymmetry. This work, at the very least, provides the baseline response for a convection-less system behavior.

### 3.8 Conclusions

This work addresses three important aspects of the $i-V$ response of a micro-nano-fluidic system subject to combined potential and diffusivity (viscosity) gradients.

First, it addresses an open question of whether or not the $i-V$ should cross the origin. We show that for any arbitrary diffusion field, $D(x)$, the $i-V$ always crosses the origin. Our continuum-based model incorporates the key components of the statistical models (i.e., the electrochemical potential) and is internally consistent in that it accounts for electrostatic interactions. Further, in contrast to the statistical mechanics models, our model is entirely consistent with experimental measurements – this provides additional substantiation to our model.

The second aspect is that the slope at which the $i-V$ crosses the origin, the Ohmic conductance, is accurately captured by our model. Not only do we have excellent correspondence to the non-approximated numerical simulations but our new model is able to recapitulate several models [Eq. (15)-(18)] when the diffusion field is uniform everywhere, and it can extend these models when the diffusion is spatially dependent. That our new model is able to recapitulate and extend existing models, also suggests that this is the correct model.

The third and final aspect of this work addresses is related to the asymmetric response of the $i-V$ leading to current rectification. To date, the rectification has been attributed to EOF and asymmetric geometries. Here, we have shown that convection-less response alone (i.e., no EOF) is able to predict the current rectification.

Thus, the results of this work are immensely important in that they address open questions and further extend existing models for the effects of diffusion (viscosity) gradients across permselective systems.

**Acknowledgments.** This work was supported by Israel Science Foundation grants 337/20 and 1953/20. We acknowledge the support of the Ilse Katz Institute for Nanoscale Science & Technology and the Pearlstone Center for Aeronautical Engineering Studies.